\documentclass[twocolumn,showpacs,aps,prb,reprint,amsfonts,amsmath,amssymb,floatfix,superscriptaddress, longbibliography]{revtex4-2}

\usepackage{graphics}
\usepackage{siunitx}
\usepackage{color}
\usepackage{hhline}
\usepackage{mathrsfs}
\usepackage{adjustbox}
\usepackage{dcolumn}
\usepackage{bm}
\usepackage{multirow}
\usepackage{booktabs}
\usepackage{afterpage}
\usepackage{amsmath}
\usepackage{physics}

\usepackage[normalem]{ulem}
\usepackage{here} 
\usepackage[inline]{asymptote}
 \usepackage{caption}
\usepackage[subrefformat=parens]{subcaption}
\usepackage[version=4]{mhchem}
\usepackage[nohyperlinks,nolist]{acronym} 
\usepackage{multirow}

\usepackage{caption} 
\usepackage{comment}
\usepackage[colorlinks=true,citecolor=blue,linkcolor=blue,urlcolor=blue]{hyperref}
\usepackage{cleveref}
\crefname{equation}{Eq.}{Eq.}
\crefname{figure}{Fig.}{Fig.}
\crefname{table}{Table}{Table}
\crefname{section}{Sec.}{Sec.}\crefname{appendix}{Appendix}{Appendix}
\Crefname{equation}{Equation}{Equation}\Crefname{figure}{Figure}{Figure}\Crefname{table}{Table}{Table}\Crefname{section}{Section}{Section}\Crefname{appendix}{Appendix}{Appendix}

\begin{document}

 \title{Transferability of the chemical bond-based machine learning model for dipole moment: the GHz to THz dielectric properties of liquid propylene glycol and polypropylene glycol}

\author{Tomohito Amano}
\affiliation{Department of Physics, The University of Tokyo, Hongo, Bunkyo-ku, Tokyo 113-0033, Japan}
\author{Tamio Yamazaki}
\affiliation{JSR-UTokyo Collaboration Hub, CURIE, JSR Corporation, 1-9-2, Higashi-Shinbashi, Minato-ku, Tokyo 105-8640, Japan}

\author{Naoki Matsumura}
\affiliation{Fujitsu Research, Fujitsu Limited, 4-1-1, Kamiodanaka, Nakahara-ku, Kawasaki, Kanagawa 211-8588, Japan}

\author{Yuta Yoshimoto}
\affiliation{Fujitsu Research, Fujitsu Limited, 4-1-1, Kamiodanaka, Nakahara-ku, Kawasaki, Kanagawa 211-8588, Japan}

\author{Shinji Tsuneyuki}
\affiliation{Department of Physics, The University of Tokyo, Hongo, Bunkyo-ku, Tokyo 113-0033, Japan}
\affiliation{Institute for Physics of Intelligence, The University of Tokyo, Hongo, Bunkyo-ku, Tokyo 113-0033, Japan}

 \date{\today}

 \onecolumngrid

\begin{abstract}
We conducted a first-principles study of the dielectric properties of liquid \ac{PG} and \ac{PPG} using a recently developed chemical bond-based \ac{ML} model for dipole moments [T. Amano et al. Phys. Rev. B 110, 165159 (2024)]. The \ac{ML} dipole models successfully predict the dipole moment of various liquid configurations in close agreement with DFT calculations and generate $\SI{20}{\nano\second}$ quantum-accuracy dipole moment trajectories to calculate the dielectric function, when combined with \ac{ML} potentials. The calculated dielectric function of \ac{PG} closely matches experimental results. We identified a libration peak at $\SI{600}{\per\cm}$ and an intermolecular mode at $\SI{100}{\per\cm}$, previously noted experimentally. Furthermore, the models trained on PG2 training data can apply to longer chain \acp{PPG} not included in the training data. The present research marks the first step toward developing a universal bond-based dipole model.
\end{abstract}

\twocolumngrid
 
\maketitle

\begin{acronym}[PPG725]
\setlength{\itemsep}{4.0pt}
     \acro  {H-bond}[H-bond] {hydrogen bond} 
     \acrodefplural{H-bond}[H-bonds] {hydrogen bonds} \acro  {DPMD}  [DPMD]   {deep potential molecular dynamics}
     \acro  {RMSE}  [RMSE]   {root mean square error}
     \acro  {RDF}   [RDF]    {radial distribution function}
     \acro  {KS}    [KS]    {Kohn-Sham} \acro  {O-lp}  [O-lp]  {O lone-pair} \acro  {DFPT}  [DFPT]  {density functional perturbation theory} \acro  {DFT}   [DFT]   {density functional theory}
     \acro  {ML}    [ML]    {machine learning}
     \acro  {VACF}  [VACF]  {velocity auto-correlation function}
     \acro  {VDOS}  [VDOS]  {vibrational density of states}
     \acro  {PG}    [PG]    {propylene glycol}
     \acro  {PG2}   [PG2]   {di-propylene glycol}
     \acro  {PPG}   [PPG]   {polypropylene glycol}
     \acro  {PPG725}[PPG725]{14-mer of Propylene glycol }
     \acro  {WF}    [WF]    {Wannier function}
     \acro  {MLWF}  [MLWF]  {maximally-localized Wannier function}
     \acrodefplural{MLWF} [MLWFs]  {maximally-localized Wannier functions}
     \acro  {WC}    [WC]    {Wannier center}
     \acrodefplural{WC} [WCs]  {Wannier centers}
     \acro  {BC}    [BC]    {bond center}
     \acro  {CPMD}  [CPMD]  {Car-Parrinello molecular dynamics}
     \acro  {BOMD}  [BOMD]  {Born-Oppenheimer molecular dynamics}
     \acro  {CMD}   [CMD]  {classical molecular dynamics}
     \acro  {BO}    [BO]    {Born-Oppenheimer}
     \acro  {AIMD}  [AIMD]  {\textit{ab initio} molecular dynamics}
     \acro  {MD}    [MD]    {molecular dynamics}
     \acro  {lp}    [lp]    {lone-pair}   
     \acro  {THz}   [THz]    {terahertz}
     \acro  {IR}    [IR]     {infrared}
\end{acronym} 

\section{Introduction}\label{sec:intro}

Computational simulations of dielectric functions are essential for understanding the dynamical, structural, and optical properties of materials. The first-principles anharmonic phonon method~\cite{fugallo2018Infrared,amano2023Lattice} is used to calculate the lattice dielectric properties of crystalline systems, whereas \ac{MD} simulations are commonly employed for liquid substances~\cite{kulschewski2013Molecular,mishra2023Computational}. The dielectric function, which is derived from the dipole autocorrelation function along \ac{MD} trajectories, requires both accurate trajectory data and precise estimation of the dipole moments of the system. In classical \ac{MD} simulations, empirical fixed charges are typically used to calculate the dipole moment. However, this approach does not account for local atomic polarization effects, sometimes making it challenging to quantitatively reproduce experimental results~\cite{carlson2020Exploring,holzl2021Dielectric,ryu2024Understanding}. The classical polarizable force fields and first-principles molecular dynamics are known methods for incorporating the effects of polarization due to many-body atomic interactions. Notable examples of the former include the Drude oscillator model~\cite{lemkul2016Empirical} and the induced dipole model~\cite{cieplak2009Polarization}, which have been successfully applied to the calculation of dielectric constants~\cite{mishra2023Computational} and dielectric functions~\cite{binninger2023AMOEBA,sharma2023Terahertz} of molecular systems.

The \ac{AIMD} simulations~\cite{car1985Unified}, which incorporate many-body atomic interactions based on quantum electronic structure calculations, have been extensively used to study the dielectric properties of various materials~\cite{handgraaf2003Initio,handgraaf2004Densityfunctional,wang2017Initio}. A key advantage of the \ac{AIMD} method is that the electronic polarization effects can be directly analyzed in a quantum mechanical manner through the \ac{MLWF} method~\cite{king-smith1993Theory,marzari1997Maximally,marzari2012Maximally}. Several \ac{AIMD} studies on molecular liquids have revealed that the polarization of the \acp{WF}, caused by the local atomic interactions, enhances the dipole moments compared to the gas phase~\cite{handgraaf2003Initio,sieffert2013Liquid} and has a significant effect on the absorption spectrum in the \ac{THz} to \ac{IR} regions~\cite{carlson2020Exploring}. Unfortunately, the high computational cost of \ac{AIMD} restricts its typical applications to systems with only hundreds of atoms and a time scale of $\SI{100}{\pico\second}$, which is insufficient to study the dielectric properties of large molecules.

To achieve the accuracy of first-principles calculations and the efficiency of empirical force fields, various \ac{ML} force fields have been actively developed~\cite{deringer2021Gaussian,behler2007Generalized,wang2018DeePMDkit,gilmer2017Neural,musaelian2023Learning}. Alongside \ac{ML} potentials, \ac{ML} models for atomic partial charges have also been developed to study dielectric properties~\cite{gastegger2017Machine,veit2020Predicting,sifain2018Discovering,beckmann2022Infrared} and to model electrostatic potentials~\cite{metcalf2021ElectronPassing,thurlemann2022Learning,ko2021Fourthgeneration,bleiziffer2018Machine,bereau2018Noncovalent,yao2018TensorMol01,unke2019PhysNet}. Unlike empirical fixed charges, \ac{ML} partial charges dynamically change their values according to the surrounding atomic environment, thereby more accurately capturing local structural features. 

A different approach, in line with the modern theory of polarization~\cite{king-smith1993Theory}, is to directly predict \acp{WC}, which represent the mass center of \acp{WF}~\cite{krishnamoorthy2021Dielectric,zhang2020Deep,sommers2020Raman,zhang2022Deep,ryu2024Understanding}. Several authors introduced the Wannier centroid, defined as the average position of the \acp{WC} in a molecule, and constructed the \ac{ML} models of the centroid to investigate the dielectric properties of \ce{H2O}~\cite{krishnamoorthy2021Dielectric,zhang2020Deep,sommers2020Raman,zhang2022Deep}. The centroid approach was also successfully applied to study the dielectric properties of crystalline systems including \ce{BaPbO3}~\cite{xie2022Initio} and \ce{KH2PO4}~\cite{yang2024Deuteration}. Furthermore, \ac{ML} molecular dipole moments methods have been developed by combining Wannier centroids and ionic charges~\cite{kapil2020Inexpensive,hou2020Dielectric}. Additionally, \ac{ML}-predicted \acp{WC} play an important role in describing long-range interactions in \ac{ML} force fields~\cite{gao2022Selfconsistent,zhang2022Deep,cools-ceuppens2022Modeling}.

However, methods that predict centroids or molecular dipoles often require larger cutoffs for \ac{ML} descriptors when applied to molecules with higher molecular weights. To address this challenge, we recently developed a chemical bond-based \ac{ML} model for \ac{WC}~\cite{amano2024Chemical}, which assigns \acp{WC} to chemical bonds and predicts the \ac{WC} for each chemical bond rather than the entire molecule. This method enables accurate dipole moment predictions without compromising precision, even for large molecules. Since the positions of the \acp{WC} do not vary significantly between different molecular species, the bond-based \ac{ML} dipole models are potentially transferable across various materials. As the previous study only reported the results for methanol and ethanol, it is desirable to investigate the transferability of the proposed \ac{ML} scheme in application to more complex molecules. Moreover, it is essential to explore the dielectric properties in frequency ranges that are inaccessible to \ac{AIMD} by combining this method with \ac{ML} potentials, which was not addressed in the previous study.

\begin{figure}[htb]  \captionsetup[subfigure]{font={bf,large}, skip=1pt, margin=-0.7cm,justification=raggedright, singlelinecheck=false}
\centering
\includegraphics[]{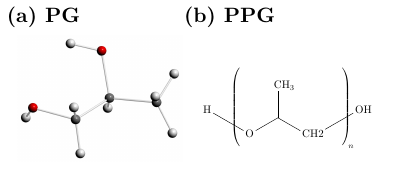}
\caption{(a) The molecular structure of \ac{PG}. The red, gray, and white spheres represent oxygen, carbon, and hydrogen atoms, respectively. (b) SMILES of \ac{PPG}, with $n=1$ corresponding to \ac{PG}. In this work, $n=2$ to $12$ were used.}
\label{fig:smiles}
\end{figure}

We selected \ac{PG}, also known as 1,2-propanediol, along with its oligomers, \ac{PPG}, to validate the transferability of our bond-based \ac{ML} approach. \ac{PG}, whose structure is depicted in \cref{fig:smiles}, is a well-known glass-forming liquid with a glass transition temperature of approximately $\SI{170}{\kelvin}$~\cite{defrancesco2002Broadband,kohler2010Glassy}. The dielectric properties in both the liquid and glass phases are extensively examined through various experiments~\cite{ikada1986Liquid,park1998Dielectric,usacheva2010Dielectric}. \ac{PPG} has also been thoroughly studied for its dielectric properties, including dimer (PG2)~\cite{kao2007Static}, trimer (PG3)~\cite{pawlus2005Hydrogen,chalikwar2017Dielectric}, PPG425 (PG7)~\cite{singh2023Dielectric,capaccioli2002Influence}, PPG725 (PG12)~\cite{koda2016Broadband}, and larger polymers~\cite{baur2004Dielectric,sengwa2002Study}. The notation $\mathrm{PG}n$ refers to the corresponding $n$-mer of PG and PPG725 indicates a molecular weight of $725$. Although many theoretical studies on \ac{PG} have relied on classical \ac{MD} simulations~\cite{ferreira2017New,brodin1998Raman,ahlstrom1998Lowfrequency,kulschewski2013Molecular}, challenges remain in the accurate prediction of dielectric properties. The optimized potential for liquid simulations (OPLS) united-atom force field~\cite{jorgensen1996Development} predicts a dielectric constant of $12$ at room temperature~\cite{kulschewski2013Molecular}, significantly underestimating the experimental value of $28$~\cite{sengwa2003Comparative,vishwam2017Dielectric}. A specialized force field for \ac{PG}~\cite{ferreira2017New} has not been used to investigate the dielectric properties. There is a strong demand for studies that utilize accurate first-principles methods to fully explore the dielectric properties of PG.

In this paper, we apply the bond-based \ac{ML} dipole approach to \ac{PG} and \acp{PPG}, demonstrating that this method can generate long-time first-principles dipole trajectories when combined with \ac{ML} potentials. Furthermore, we show that the model trained on PG2 data can be directly applied to higher molecular weight \acp{PPG} without additional fine-tuning. To perform \ac{MD} simulations with first-principles accuracy, we trained \ac{ML} potentials for \ac{PG} and \ac{PG2} using the DeepMD-kit~\cite{wang2018DeePMDkit,zeng2023DeePMDkit}. We first studied the dielectric properties of \ac{PG} using the trained \ac{ML} potentials and the \ac{ML} dipole models. To the best of our knowledge, this is the first detailed \textit{ab initio} study on the dielectric properties of \ac{PG}. The calculated dielectric spectrum closely matches the experimental data across a wide range of frequencies from GHz to THz regions, revealing a hydroxyl hydrogen libration peak at $\SI{600}{\per\cm}$ and a mode at $\SI{100}{\per\cm}$ attributed to intermolecular interactions. Subsequently, we employed the \ac{ML} potentials and \ac{ML} dipoles trained on PG2 to predict the dipole moment and dielectric function of higher molecular weight \acp{PPG}. Our calculations closely reproduced the experimental results, demonstrating the transferability of our \ac{ML} scheme.

\section{Theory}\label{sec:theory}
 \begin{figure}[htb]\centering
\captionsetup[subfigure]{font={bf,large}, skip=1pt, margin=-0.7cm,justification=raggedright, singlelinecheck=false}
 \begin{subcaptionblock}{0.45\linewidth}
\adjustbox{right}{\includegraphics[width=\textwidth]{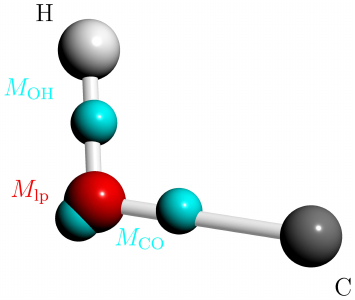}}
  \end{subcaptionblock}\hfill
  \begin{subcaptionblock}{0.45\linewidth}
\adjustbox{right}{\includegraphics[width=\textwidth]{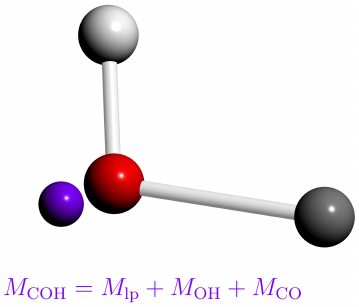}}
 \end{subcaptionblock}
 \caption{The schematic images of the \ce{COH} bond dipole. (left) The carbon (gray), oxygen (red), and hydrogen (white) atoms constituting the \ce{COH} bond with three corresponding \acp{WC} (cyan). (right) The \ce{COH} bond Wannier center (purple).}
\label{fig:coh_bond}
 \end{figure} 

\subsection{Dipole Moments}

The \ac{MLWF} method decomposes the total dipole moment of the system into the ionic and the electronic parts as~\cite{marzari1997Maximally,marzari2012Maximally}
\begin{align}\label{EqWF}
 \vb{M} = e\sum_{i=1}^{N} Z_{i}\vb{r}_{i} -2e\sum_{k=1}^{N_w}\vb{w}_{k},
\end{align}
where $Z_{i}$ and $\vb{r}_i$ are the ionic charge and nuclear position of the $i$-th atom, respectively. $\vb{w}_{k}$ describes the position of the $k$-th \ac{WC}. $N$ and $N_{w}$ represent the number of atoms and \acp{WC}, respectively. The factor $2$ in the second term represents the spin degrees of freedom. We employ a valence-only pseudopotential method, where the nuclear charges $eZ_{i}$ encompass contributions from both the nuclei and the frozen core electrons, while the \acp{WC} correspond to the valence electrons.

Following the bond dipole scheme proposed in Ref.~\cite{amano2024Chemical}, we assign each \ac{WC} to oxygen lone pairs or chemical bonds. For simplicity, We consider a system consisting of carbons, oxygens, and hydrogens with single bonds. The total dipole moment in \cref{EqWF} can be rewritten in the bond-based form as 
\begin{align}
 &\vb{M} = \sum_{i=1}^{N_\mathrm{single}}\vb*{\mu}_{i}^{\mathrm{single}}+\sum_{i=1}^{N_\mathrm{lp}}\vb*{\mu}_{i}^{\mathrm{lp}}\label{Eq:bd} \\
 &\vb*{\mu}_{i}^{\mathrm{single}} = 2e\left(\vb{r}^{s}_{i}-\vb{w}^{\mathrm{s}}_{i}\right) \\
 &\vb*{\mu}_{i}^{\mathrm{lp}}     = 4e\left(\vb{r}^{\ce{O}}_{i}-\vb{w}^{\mathrm{lp}}_{i}\right),
\end{align}
where $\vb{r}^{s}_{i}$ and $\vb{r}^{\ce{O}}_{i}$ are the positions of the $i$-th \ac{BC} of the chemical bonds and oxygens, and $\vb{w}^{\mathrm{s}}_{i}$ and $\vb{w}^{\mathrm{lp}}_{i}$ are the positions of assigned \acp{WC}. $N_{\mathrm{single}}$ and $N_{\mathrm{lp}}$ are the number of single bonds and lone pairs, respectively. We refer to $\vb*{\mu}_{i}^{\mathrm{single}}$ and $\vb*{\mu}_{i}^{\mathrm{lp}}$ as bond dipoles, which are simply the relative vectors from the \acp{WC} to the \acp{BC} or oxygens. 

\Cref{Eq:bd} decomposes the dipole moment into its most fundamental components, allowing for the definition of a new virtual bond dipole by summing the contributions from arbitrary bond dipoles. Since the oxygen lone pair is thought to correlate with the \acp{WC} of the adjacent bonds, combining it with surrounding bonds would achieve both maintaining accuracy and decreasing computational cost. We define a new bond dipole by aggregating the contributions from the lone pair $i$ and the two adjacent bonds $a(i)$ and $b(i)$. In systems consisting only of carbons, hydrogens, and oxygens, two distinct bond dipoles can be defined, $\vb*{\mu}^{\mathrm{COH}}$ and $\vb*{\mu}^{\mathrm{COC}}$:

\begin{align}
 &\vb*{\mu}_{i}^{\mathrm{COH}} = \vb*{\mu}_{a(i)}^{\mathrm{CO}} +  \vb*{\mu}_{i}^{\mathrm{lp}} + \vb*{\mu}_{b(i)}^{\mathrm{OH}} \\
 &\vb*{\mu}_{i}^{\mathrm{COC}} = \vb*{\mu}_{a(i)}^{\mathrm{CO}} +  \vb*{\mu}_{i}^{\mathrm{lp}} + \vb*{\mu}_{b(i)}^{\mathrm{CO}}.
\end{align}
\Cref{fig:coh_bond} illustrates the concept of the \ce{COH} dipole moment. The total dipole moment in \cref{Eq:bd} is rewritten in terms of \ce{CH}, \ce{CC}, \ce{COC}, and \ce{COH} bond dipole moments as
\begin{align}\label{eq:dipole}
 \vb{M}  &=\sum_{i=1}^{N_\mathrm{CH}}\vb*{\mu}_{i}^{\mathrm{CH}}+\sum_{i=1}^{N_\mathrm{CC}}\vb*{\mu}_{i}^{\mathrm{CC}}+\sum_{i=1}^{N_\mathrm{COC}}\vb*{\mu}_{i}^{\mathrm{COC}}+\sum_{i=1}^{N_\mathrm{COH}}\vb*{\mu}_{i}^{\mathrm{COH}},
\end{align}
where $N_\mathrm{CH}$, $N_\mathrm{CC}$, $N_\mathrm{COC}$, and $N_\mathrm{COH}$ are the numbers of each bond.

\subsection{The dielectric properties}

The dielectric function $\varepsilon(\omega)$ can be calculated by the Fourier transform of the derivative of the autocorrelation function of the dipole moment as~\cite{neumann1984Computer}
\begin{align}
\varepsilon(\omega)&= \varepsilon'(\omega)-i\varepsilon''(\omega) \label{Eq:diel1} \\
 &=\varepsilon^{\infty} -\frac{1}{3\varepsilon^{0}k_{\mathrm{B}}TV} \int_{0}^{\infty}\left(\frac{\dd \expval{\vb{M}(0)\cdot \vb{M}(t)}}{\dd t}\right)e^{-i\omega t}\dd t. \label{Eq:diel2}
\end{align}
where $\varepsilon^{\infty}$ is the high-frequency dielectric constant, $k_{\mathrm{B}}$ is the Boltzmann constant, $T$ is the temperature, $V$ is the volume of the simulation cell, $\varepsilon^{0}$ is the dielectric constant in vacuum, and $\vb{M}$ represents the total dipole moment of the system. $\expval{}$ denotes the canonical ensemble average. $\varepsilon'(\omega)$ and $\varepsilon''(\omega)$ represent the real and imaginary part of the dielectric function, respectively. To avoid evaluating the derivative of the auto-correlation function, we adopt the alternative form derived through the integration by parts of \cref{Eq:diel2}~\cite{cardona2018Molecular}
 \begin{align}
\varepsilon(\omega)
 &=\varepsilon^{\infty} -\frac{i\omega}{3\varepsilon^{0}k_{\mathrm{B}}TV} \int_{0}^{\infty}\expval{\vb{M}(0)\cdot \vb{M}(t)}e^{-i\omega t}\dd t. \\
&=\varepsilon^{\infty} +\frac{1}{3\varepsilon^{0}k_{\mathrm{B}}TV i\omega} \int_{0}^{\infty}\expval{\vb{\dot{M}}(0)\cdot \vb{\dot{M}}(t)} e^{-i\omega t}\dd t.
 \end{align}
The imaginary part of the dielectric function is explicitly given as~\cite{wang2017Initio}
\begin{align}
 \varepsilon''(\omega) &= \frac{\omega}{3\varepsilon^{0}k_{\mathrm{B}}TV} \Re\int_{0}^{\infty}\expval{\vb{M}(0)\cdot \vb{M}(t)}e^{-i\omega t}\dd t \\
&=\frac{\omega}{6\varepsilon^{0}k_{\mathrm{B}}TV} \int_{-\infty}^{\infty}\expval{\vb{M}(0)\cdot \vb{M}(t)}e^{-i\omega t}\dd t.  
\end{align}
The absorption coefficient per unit length $\alpha(\omega)$ is defined through Lambert-Beer's law~\cite{hecht2017Optics}: \begin{align}
\alpha(\omega) 
       &= \frac{\omega}{c}\frac{\varepsilon''(\omega)}{n(\omega)} \\ 
       &= \frac{\omega^2}{6\varepsilon^{0}cn(\omega)k_{\mathrm{B}}TV} \int_{-\infty}^{\infty}\expval{\vb{M}(0)\cdot \vb{M}(t)}e^{-i\omega t}\dd t.  
 \end{align}
 where $c$ is the speed of light in vacuum, and $n(\omega)$ is the refractive index calculated from the dielectric function. Different coefficients are sometimes adopted in the literature~\cite{sharma2005Intermolecular,ramirez2004Quantum}, depending on the classical approximation applied to the Kubo formula in \cref{Eq:diel2}.

Similarly, we define the power spectrum $I_x(\omega)$ of a real-valued time series $x(t)$ as the Fourier transform of the normalized time autocorrelation function:
\begin{align}\label{Eq:power}
  I_{x}(\omega) = \int_{-\infty}^{\infty} \frac{\expval{x(0)x(t)}}{\expval{x(0)^2}}e^{-i\omega t}\dd t.
\end{align}
If $x(t)$ is the vector quantity, one may evaluate inner dot of $x(t)$ as 
\begin{align}
  I_{\vb{x}}(\omega) = 
  \int_{-\infty}^{\infty} \frac{\expval{\vb{x}(0)\cdot\vb{x}(t)}}{\expval{\left|\vb{x}(0)\right|^2}}e^{-i\omega t}\dd t.
\end{align}
A detailed analysis of MD trajectories becomes possible by considering various physical quantities, such as interatomic distances and angles for $x$~\cite{carlson2020Exploring,liu2022Highlevel}. One good example is the \ac{VDOS}, which of the $i$-th atom is given by the Fourier transform of the \ac{VACF} as
\begin{align}\label{Eq:vdos}
 D_{i}(\omega) = \int_{-\infty}^{\infty} \expval{\vb*{v}_{i}(0)\cdot\vb*{v}_{i}(t)}e^{-i\omega t}\dd t,
\end{align}
where $\vb*{v}_{i}(t)$ is the atomic velocity at time $t$. The total \ac{VDOS} is given by averaging over all atomic contribution of \cref{Eq:vdos}:
\begin{align}
 D(\omega) = \frac{1}{N}\sum_{i=1}^{N}D_{i}(\omega).
\end{align}
We can calculate the \ac{VACF} for each atomic species by restricting the summation. 

In practice, the time integration is evaluated using the convolution of the Fourier transform. Using the Fourier transformed atomic velocity
\begin{align}
 \vb*{v}_{i}(\omega)= \int_{-\infty}^{\infty}\vb*{v}_{i}(t)e^{-i\omega t} \dd t, 
\end{align}
the atomic \ac{VDOS} is obtained by 
\begin{align}
 D_{i}(\omega) = \vb*{v}^{*}_{i}(\omega)\cdot\vb*{v}_{i}(\omega),
\end{align}
which is considerably faster than computing \cref{Eq:vdos} directly.

\subsection{Machine learning model}
We follow the neural network architecture proposed by Zhang et al.~\cite{zhang2020Deep}, which is suitable for predicting vector quantities by considering rotational symmetry. We consider a \ac{ML} model that predicts the $i$-th bond dipole $\vb*{\mu}_i$ based on the input $\vb{r}_k\in \mathscr{N}_{i}$, where $\mathscr{N}_i$ represents the atoms located within the cutoff radius $r_{c}$ around the \ac{BC} or oxygen atom $i$, and $N_{i}$  denotes the total number of such atoms. In this approach, the atomic environments are embedded in a rotationally invariant feature matrix through a first neural network, named the embedding network, and the feature matrix is used as an input of the second neural network, called the fitting network, to predict the bond dipole moment. 

Hereafter, we use a coordinate system where the origin is set at the position of the $i$-th \ac{BC} or oxygen atom, and the coordinates of the $k$-th atom in this system are denoted as $\vb{r}_{ik}$. The following four-component vectors represent the atomic coordinates:
\begin{align}
\vb{q}_{ik}&= (q^{1}_{ik},q^{2}_{ik},q^{3}_{ik},q^{4}_{ik}) \\
           &= s(r_{ik})(1,x_{ik},y_{ik},z_{ik}),
\end{align}
where $s(r)$ is the cutoff function that is equal to $1/r'_{ki}$ in $r'_{ki}<r_{\mathrm{c}_{0}}$ and decays to zero as $r'_{ki}$ approaches $r_{\mathrm{c}}$. The $N_i$ by $4$ matrix $\vb{Q}_{i}=(Q_{k\lambda})=(q^{\lambda}_{ik})$, which is explicitly written as 
\begin{align}
Q_{i}&= 
    \begin{pmatrix} 
        q^1_{i1},q^2_{i1},q^3_{i1},q^3_{i1} \\
        \cdots\\
        q^1_{iN_i},q^2_{iN_i},q^3_{iN_i},q^3_{iN_i} 
    \end{pmatrix},
\end{align}
contains information about the surrounding atoms of the $i$-th \ac{BC} or oxygen atom. The atomic index $k=1,2,\cdots, N_i$ is sorted first based on the type of neighboring atoms, and then on the atomic distances $q^1_{ik}$. For each atomic type, the number up to a specified maximum number of atoms is considered. To keep the length of the descriptor constant, the descriptors with different atomic types are padded with zero if the neighbor lists of atomic species are smaller than the specified value. The embedding neural network takes only the distance information $\{q^{1}_{ik} | k=1,\cdots,N_{i}\}$ from $Q_i$ as input and produces an output with $MN_i$ elements, represented by the $M\times N_i$ matrix $E$, where $M$ is the hyperparameter. Noting the first $M'(<M)$ rows of $E$ as a new matrix $E'$, we define the feature matrix $D_i$ as 
\begin{align}\label{Eq:feature}
 \vb{D}_{i} = (\vb{E}\vb{Q}_{i})(\vb{E}'\vb{Q}_{i})^{T} = \vb{E}\vb{Q}_{i}\vb{Q}^{T}_{i}\vb{E}'^{T},
\end{align}
which is a $M\times M'$ matrix.

The fitting network takes $D_i$ as input and produces $M$ outputs $F_{j} (j=1,2,\cdots, M)$, which are ultimately used to compute the bond dipoles $\vb*{\mu}_{i}=(\mu^{1}_{i},\mu^{2}_{i},\mu^{3}_{i})$ with the last three columns of $\vb{T}=\vb{E}\vb{Q}$,
\begin{align}
 \mu_{i}^{\lambda} = \sum_{j=1}^{M} F_j(\vb{D}) T_{j,\lambda+1} \, (\lambda=1,2,3).
\end{align}

\section{Computational details}

\begin{figure}[thb]\centering
\includegraphics[]{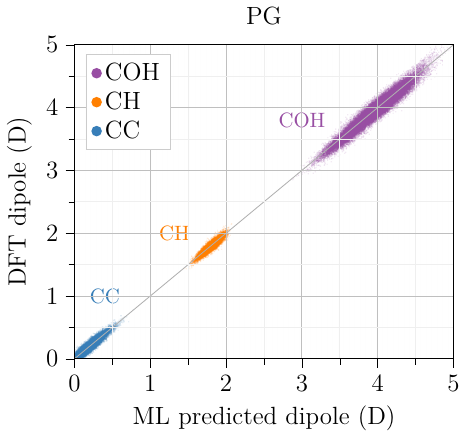}
\caption{Learning accuracy of \ac{ML} bond dipoles of \ac{PG}. The blue, orange, and purple dots represent the absolute values of the CC, CH, and COH bond dipoles, respectively.}
 \label{fig:lr_pg}
 \end{figure} 

\begin{table}[tb] \centering
\caption{RMSE $[\mathrm{D}]$ of \ac{ML} dipole models for \ac{PG} and PG2. The \ce{CO}, \ce{O}, and \ce{OH} models were trained for reference purposes and were not used for spectrum calculations. The CO+Olp+OH and CO+Olp+CO represents the COH and COC dipoles calculated from these three models, respectively.}
 \renewcommand{\arraystretch}{1.5}
{\tabcolsep = 0.15cm
  \begin{tabular}{lcc}
   \hline\hline
     RMSE $[D]$ & PG & PG2 \\ \hline
     CC         &  0.038  & 0.033 \\ 
     CH         &  0.031  & 0.027 \\ 
     COH        &  0.070  & 0.066 \\ 
     COC        &  -      & 0.077 \\ \hline
     CO         &  0.031  & 0.032 \\ 
     Olp        &  0.056  & 0.052 \\ 
     OH         &  0.023  & 0.023 \\
     CO+Olp+OH  &  0.072  & 0.071 \\ 
     CO+Olp+CO  &  -      & 0.080 \\ 
     \hline\hline 

  \end{tabular}
}
\renewcommand{\arraystretch}{1.0}

 \label{table:rmse}
\end{table} \begin{figure}[htb]  \captionsetup[subfigure]{font={bf,large}, skip=1pt, margin=-0.7cm,justification=raggedright, singlelinecheck=false}
\centering
\begin{subcaptionblock}{\linewidth}
\includegraphics[]{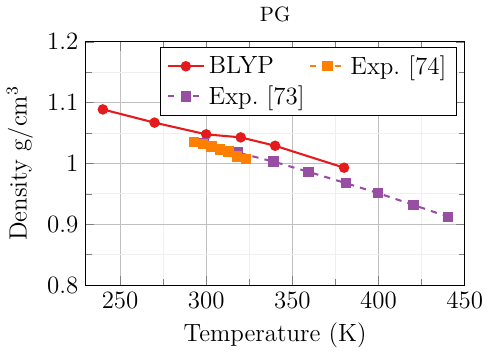}
 \end{subcaptionblock}\hfill
\caption{Calculated density of PG from $\SI{240}{\kelvin}$ to $\SI{420}{\kelvin}$ with experimental values from Sun et al.~\cite{sun2004Density} and Khattab et al.~\cite{khattab2017Density}.}
\label{fig:density}
\end{figure} \begin{figure}[htb]  \captionsetup[subfigure]{font={bf,large}, skip=1pt, margin=-0.7cm,justification=raggedright, singlelinecheck=false}
\centering
\begin{subcaptionblock}{\linewidth}
\includegraphics[]{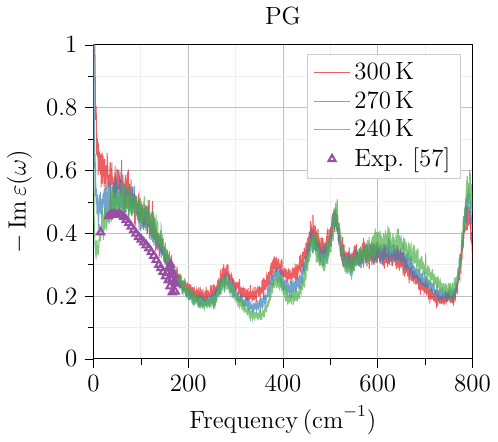}
 \end{subcaptionblock}
\caption{Calculated imaginary part of dielectric functions of \ac{PG} at three different temperatures: $240$, $270$, and $\SI{300}{\kelvin}$ with experimental values at $\SI{300}{\kelvin}$~\cite{koda2016Broadband}. The spectra were calculated from five independent $\SI{2}{\nano\second}$ trajectories for each temperature, and smoothed using the moving average method.}
\label{fig:dielec_pg}
\end{figure} \begin{figure}[htb]  \captionsetup[subfigure]{font={bf,large}, skip=1pt, margin=-0.7cm,justification=raggedright, singlelinecheck=false}
\centering
\begin{subcaptionblock}{\linewidth}
\centering
\subcaption{}\label{fig:vdosa}
\includegraphics[]{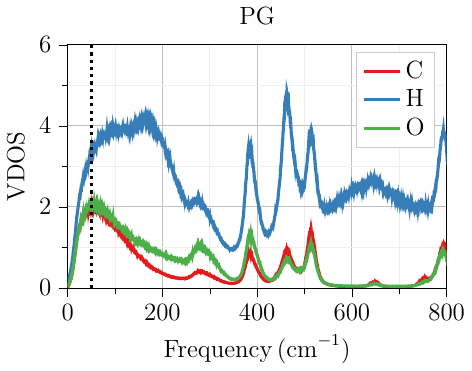} \end{subcaptionblock}
\begin{subcaptionblock}{\linewidth}
\centering
\subcaption{}\label{fig:vdosb}
\includegraphics[]{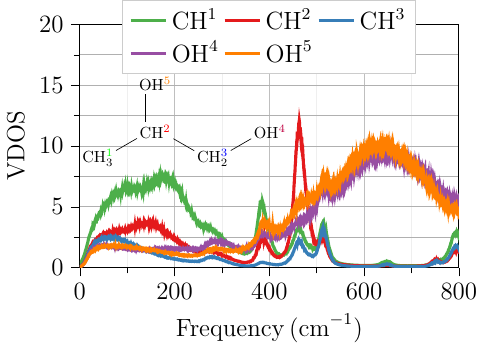} \end{subcaptionblock}
\caption{(a) Calculated \ac{VDOS} of carbon, oxygen, and hydrogen at $\SI{240}{\kelvin}$. The position of the \ac{VDOS} peak observed in the experimental dielectric function~\cite{koda2016Broadband} is indicated by the black dashed line. (b) Calculated \ac{VDOS} for hydrogen, decomposed according to its chemical environment.}
\label{fig:vdos}
\end{figure} \begin{figure}[htb]  \captionsetup[subfigure]{font={bf,large}, skip=1pt, margin=-0.7cm,justification=raggedright, singlelinecheck=false}
\centering
\begin{subcaptionblock}{\linewidth}
\includegraphics[]{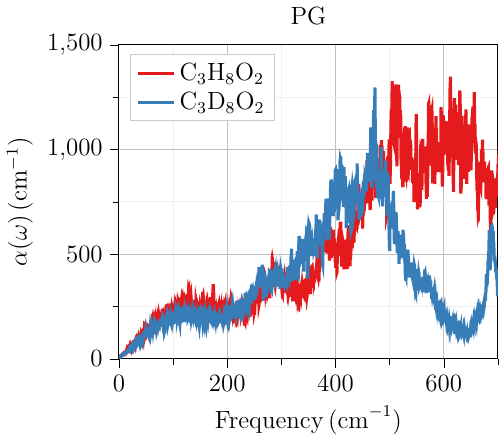} \end{subcaptionblock}\hfill
\caption{Calculated absorption spectra of \ac{PG} (red) and deuterated \ac{PG} (blue) at $\SI{300}{\kelvin}$. In the legend, \ce{H} represents normal hydrogen, while \ce{D} denotes deuterium.}
\label{fig:deutron}
\end{figure} \begin{figure*}[htb]  \captionsetup[subfigure]{font={bf,large}, skip=1pt, margin=-0.7cm,justification=raggedright, singlelinecheck=false}
\centering
\begin{subcaptionblock}{0.45\linewidth}
\subcaption{}\label{fig:dielec_pg_a}
\adjustbox{right}{\includegraphics[width=\textwidth]{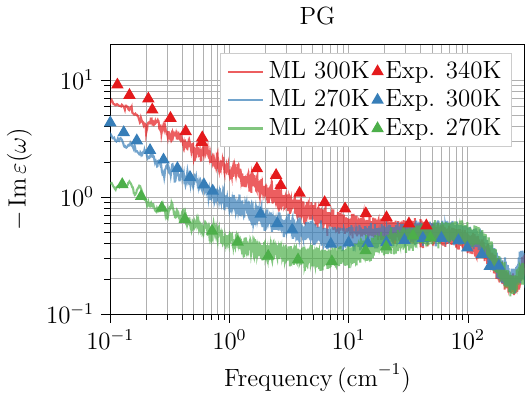}}
 \end{subcaptionblock}
\begin{subcaptionblock}{0.45\linewidth}
\subcaption{}\label{fig:dielec_pg_b}
\adjustbox{right}{\includegraphics[width=\textwidth]{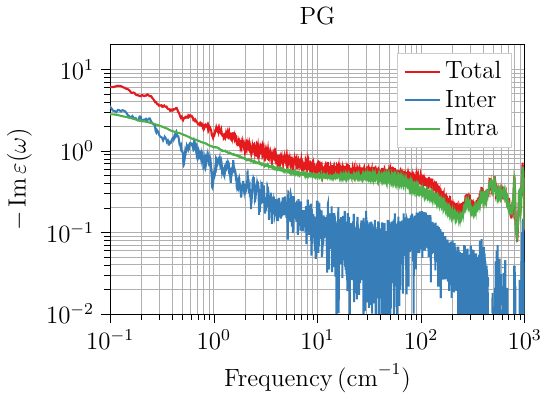}}
 \end{subcaptionblock} \hfill

\caption{(a) Calculated imaginary part of dielectric functions of \ac{PG} (lines) in logarithmic scale at three different temperature: $240$ (green), $270$ (blue), and $\SI{300}{\kelvin}$ (red). The experimental values (symbols) are taken from Kohler et al.~\cite{kohler2010Glassy} at $270$, $300$, and $\SI{340}{\kelvin}$ and Koda et al.~\cite{koda2016Broadband} at $\SI{300}{\kelvin}$. The spectra were calculated from five independent $\SI{20}{\nano\second}$ trajectories for each temperature, and smoothed using the moving average method. (b) The total (red), intramolecular (green), and intermolecular (blue) components of the calculated imaginary part of dielectric functions of \ac{PG} at $\SI{300}{\kelvin}$ according to \cref{Eq:decompose}. The intermolecular part peaks at $\SI{100}{\per\cm}$, while the intramolecular part peaks at $\SI{50}{\per\cm}$, corresponding to the \ac{VDOS} peak.}
\label{fig:dielec_pg}
\end{figure*} \begin{figure}[htb]  \captionsetup[subfigure]{font={bf,large}, skip=1pt, margin=-0.7cm,justification=raggedright, singlelinecheck=false}
\centering
\begin{subcaptionblock}{\linewidth}
\includegraphics[width=\textwidth]{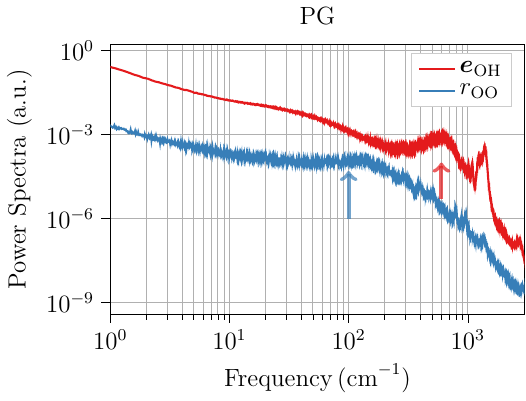}
 \end{subcaptionblock}\hfill
\caption{Calculated power spectra of $r_{\mathrm{OO}}$ and $\vb*{e}_{\mathrm{OH}}$ according to \cref{Eq:power} of PG at $\SI{240}{\kelvin}$. The peak positions of the intermolecular mode at $\SI{100}{\per\cm}$ and the librational mode at $\SI{600}{\per\cm}$ are indicated with blue and red arrows, respectively.}
\label{fig:hbond_pg}
\end{figure}

\subsection{Classical molecular dynamics simulation}
We performed the \ac{CMD} calculations using the GROMACS package~\cite{abraham2015GROMACS} to prepare initial configurations for the \ac{ML} molecular dynamics simulations. In each system, molecules were positioned in the simulation box using the Packmol package~\cite{martinez2009PACKMOL} to generate the initial configurations. The force field parameters stem from the general AMBER force field (GAFF2)~\cite{wang2004Development} and the atomic charges from the AM1-BCC charge~\cite{jakalian2002Fast}, respectively, with topology files generated by the Antechamber~\cite{wang2006Automatic} and ACPYPE package~\cite{sousadasilva2012ACPYPE}. The simulations were performed in an NVT ensemble at $\SI{300}{\kelvin}$ using velocity rescaling, and the equations of motion were integrated using the Verlet algorithm with a $\SI{1}{\femto\second}$ time step.

\subsection{\textit{Ab initio} electronic structure calculation}
We employed the CPMD package v.4.3~\cite{CarParrinello} for \textit{ab initio} electronic ground state calculations and \ac{WC} constructions to prepare \ac{ML} training data. The Becke-Lee-Yang-Parr (BLYP) functional~\cite{lee1988Development,becke1988Densityfunctional}, a family of the generalized gradient approximation (GGA) scheme, was used for the exchange-correlation functional. We included Grimme's dispersion correction D2~\cite{grimme2006Semiempirical} to account for van der Waals interactions. The valence-only Goedecker-Teter-Hutter (GTH) pseudopotentials~\cite{goedecker1996Separable} were used, and the plane waves represented the electronic wavefunctions with an energy cutoff of $\SI{100}{\mathrm{Ry}}$. The energy convergence criteria was set to $\SI{1e-7}{\hartree}$. Only the $\Gamma$-point was considered for all calculations.

\subsection{Training of the machine learning models}
We employed the DeePMD-kit package~\cite{wang2018DeePMDkit} to construct neural network many-body potentials for \ac{PG} and \ac{PG2}. The active learning scheme implemented in the GeNNIP4MD~\cite{Fujitsu} was used to generate training structures. Starting from $3000$ structures collected from \ac{AIMD} calculations at $300$, $600$, and $\SI{800}{\kelvin}$, we carried out active sampling using NPT and NVT calculations at $300$, $600$, and $\SI{800}{\kelvin}$. We also used the volume scan technique~\cite{magdau2023Machine} to increase the accuracy and robustness of the potentials. We sampled $14996$ and $27438$ structures for \ac{PG} and PG2, respectively, which were used as training data for the ML potentials. The neural network architecture includes an embedding network of three layers with $25$, $50$, and $100$ neurons, respectively, and a fitting network with a three-layer structure, each containing $100$ neurons.

We utilized the PyTorch-based MLWC (machine learning Wannier center) package~\cite{ToAmano} to assign \acp{WC} and construct the \ac{ML} models for dipole moments of \ac{PG} and \ac{PG2}. We used the same structures obtained from the active learning procedure. The descriptors contained up to $24$ neighboring atoms for each atomic species, with an outer cutoff of $\SI{6}{\angstrom}$ and an inner cutoff of $\SI{4}{\angstrom}$. The size of the feature matrix in \cref{Eq:feature} was set to $M=20$ and $M'=6$. 
For both \ac{ML} potentials and dipole moments, we used $90\%$ of the reference data for training and the remaining $10\%$ of the data points for validation.

\subsection{Deep potential molecular dynamics with machine learning dipole moment}

We used the LAMMPS package~\cite{thompson2022LAMMPS} to perform \ac{DPMD} simulations combined with \ac{ML} dipole moments for evaluating the dielectric properties of \ac{PG} and \acp{PPG}. After initial configurations were generated from $\SI{10}{\nano\second}$ \ac{CMD} simulations, we performed a $\SI{200}{\pico\second}$ equilibration, from which production runs of $2$ or $\SI{4}{\nano\second}$ and $\SI{20}{\nano\second}$ were carried out. The \acp{WC} are sampled every $2.5$ and $\SI{10}{\femto\second}$, respectively. All the \ac{DPMD} simulations were conducted in an NVT ensemble at various temperatures with a Nos\'{e}-Hoover thermostat, with the integration time step of $\SI{0.5}{\femto\second}$. Calculations were performed on $64$ molecule systems for all the materials. Densities were determined from NPT calculations of $\SI{500}{\pico\second}$.

As the estimation of $\varepsilon^{\infty}$ was not within the scope of this work, we evaluated it from the square of the refractive index $n$, using the relation $\varepsilon^{\infty}=n^2$. The experimental values~\cite{assessment2009CRC,vishwam2017Microwavea} of  $n=1.43$ for \ac{PG} and $n=1.45$ for \ac{PPG} were used.

 \section{Results and Discussion}
\subsection{Propylene glycol}

We first evaluate the accuracy of the trained \ac{ML} dipole models and potentials. \Cref{fig:lr_pg} shows that the predicted dipole moments by our \ac{ML} models align excellently with the DFT reference data for all the bond types. We randomly sampled one thousand validation structures from a $\SI{20}{\pico\second}$ \ac{AIMD} trajectory at $\SI{300}{\kelvin}$. The bond dipoles increase in the order of \ce{CC}, \ce{CH}, and \ce{COH}, with average values of $0.1$, $0.9$, and $\SI{2.0}{\mathrm{D}}$, respectively. The bond dipole moment of $\SI{1}{\mathrm{D}}$ corresponds to about $\SI{0.1}{\angstrom}$ of the displacement of \ac{WC}. In the case of the \ce{CC} bond, the \ac{WC} is located near the \ac{BC} due to symmetry, resulting in a small bond dipole. In contrast, the \ac{WC} associated with \ce{CH} bonds is located closer to the H atom than the C atom. \Cref{table:rmse} presents the \ac{RMSE} for each bond, showing that the \ce{COH} model provides slightly better accuracy than the sum of the \ce{CO}, lone pair, and \ce{OH} models. The newly conceived \ce{COH} model can achieve comparable accuracy while reducing computational cost, making it particularly advantageous for long-time and large-scale calculations. In the following calculations, we calculate dipole moments using the \ce{COH} model as described in \cref{eq:dipole}, unless otherwise stated.

\Cref{fig:density} compares the calculated densities over the temperature range of $240$ to $\SI{360}{\kelvin}$ with experimental values~\cite{sun2004Density,khattab2017Density} to assess the accuracy of the trained \ac{ML} potential. The boiling point of \ac{PG} is approximately $\SI{460}{\kelvin}$, towards which the density gradually decreases. At $\SI{300}{\kelvin}$, the simulated density of $\SI{1.05}{g/cm^3}$ closely matches the experimental density of $\SI{1.03}{g/cm^3}$~\cite{sun2004Density}. Our \ac{ML} potential accurately reproduces the experimental values across a wide range of temperatures, albeit with a slight overestimation. This is because active sampling enabled us to included diverse structures spanning a broad range of temperatures and densities in the training data.

After confirming the high accuracy of the trained \ac{ML} dipole and potential models, we studied the dielectric properties of \ac{PG}. \Cref{fig:dielec_pg} shows the calculated dielectric function in the THz to the far-IR region at three temperatures: $240$, $270$, and $\SI{300}{\kelvin}$, with experimental values at $\SI{300}{\kelvin}$~\cite{koda2016Broadband}. Each spectrum was calculated from five independent $\SI{2}{\nano\second}$ trajectories. Our calculations are in good agreement with the experimental values. The spectra are highly complex, characterized by a broad peak around $\SI{50}{\per\cm}$ and another around $\SI{600}{\per\cm}$ as the main features, with some sharp peaks between $200$ to $\SI{500}{\per\cm}$. The peak around $\SI{50}{\per\cm}$ is often referred to as the \ac{VDOS} peak, as the total \ac{VDOS} also has a peak at the same frequency~\cite{koda2016Broadband,ahlstrom1998Lowfrequency}. The \ac{VDOS} peak has been studied in relation to the boson peak in the supercooled or glassy states~\cite{schneider1998Dielectric,yomogida2010Comparative,yomogida2010Comparativea,kabeya2016Boson}. The temperature dependence of the spectra is small above $\SI{400}{\per\cm}$, while a concavity emerges in the low-frequency region with reduced temperatures, as the orientational relaxation peak in the GHz range shifts to lower frequencies at reduced temperatures~\cite{kohler2010Glassy}.

To identify the origin of the multiple peaks in the THz region, \Cref{fig:vdosa} presents the calculated \ac{VDOS} for each atomic species at $\SI{240}{\kelvin}$. The \ac{VDOS} was computed using the final $\SI{200}{\pico\second}$ of the $\SI{2}{\nano\second}$ trajectories used for calculating the dielectric function. As with the dielectric function, we observed the sharp peaks at $280$, $390$, $460$, and $\SI{520}{\per\cm}$, which are discussed in \cref{appendix:A}. While the broad peak around $\SI{600}{\per\cm}$ is primarily due to hydrogen motion, the low frequency peak around $\SI{100}{\per\cm}$ shows contributions from all the atomic species. The carbon and hydrogen peaks near $\SI{50}{\per\cm}$, called the \ac{VDOS} peak, while the hydrogen shows a broader peak up to about $\SI{180}{\per\cm}$. \Cref{fig:vdosb} further illustrates the \ac{VDOS} for different types of hydrogen, revealing that all the types of hydrogens contributed to the VDOS peak at $\SI{50}{\per\cm}$, and the alkyl hydrogens are responsible for the higher frequencies up to $\SI{200}{\per\cm}$. On the other hand, the $\SI{600}{\per\cm}$ peak is attributed to the hydroxyl hydrogens. This has bee observed in other hydrogen-bonded molecular liquids, such as water~\cite{sharma2023Terahertz} and methanol~\cite{torii2023Intermolecular}, and is ascribed to the librational (vibrational) motion of the hydroxyl hydrogen atoms. \Cref{fig:deutron} compares the absorption spectra of both deuterated and normal \ac{PG} and shows that the prominent absorption peak at $\SI{600}{\per\cm}$ in normal \ac{PG} is red-shifted to approximately $\SI{500}{\per\cm}$ in deuterated \ac{PG}, leading us to conclude that the peak is indeed due to the libration.

\Cref{fig:dielec_pg_a} displays the dielectric function in the GHz frequency range, calculated at three temperatures: $240$, $270$, and $\SI{300}{\kelvin}$, with experimental values at $270$, $300$, and $\SI{340}{\kelvin}$~\cite{kohler2010Glassy}. These results were obtained by averaging five independent $\SI{20}{\nano\second}$ \ac{MD} trajectories, which far exceeds the time scales accessible by \ac{AIMD}, highlighting the extended capabilities of the \ac{ML} potentials and dipole moments. The frequency range corresponds to the high-frequency part of the orientational relaxation peak. In the case of \ac{PG}, the orientational relaxation peak consists of a single peak; however, experimental data can be well-fitted using two Havriliak-Negami (HN) functions, with the low-frequency side referred to as the $\alpha$ process and the high-frequency side as the $\beta$ process~\cite{koda2016Broadband,grzybowska2006Changes}. The $\alpha$ process is an orientation relaxation involving multiple molecules, whereas $\beta$ processes originate from several physical origins, including the JG $\beta$ process due to intermolecular interactions~\cite{ngai2001Nature}. At $\SI{300}{\kelvin}$, the peak frequency of the alpha process, determined by fitting with the HN function, is located at $\SI{0.013}{\per\cm}\, (\SI{0.4}{\giga\hertz})$~\cite{koda2016Broadband}. Our calculations accurately reproduce the experimental behavior that the peak position shifts significantly towards lower frequencies as the temperature decreases. As mentioned, a separation between the temperature-independent VDOS peak and the temperature-dependent beta process occurs at low temperature, causing a concavity in the spectrum at around $\SI{4}{\per\cm}$. Quantitatively, the results at $\SI{270}{\kelvin}$ closely match the experimental values at $\SI{300}{\kelvin}$. This minor discrepancy likely arises from approximations involving the choice of the exchange-correlation functionals or neglecting nuclear quantum effects~\cite{xu2020Isotope,krishnamoorthy2021Dielectric}. 

\begin{figure*}[htb]\centering
 \includegraphics[]{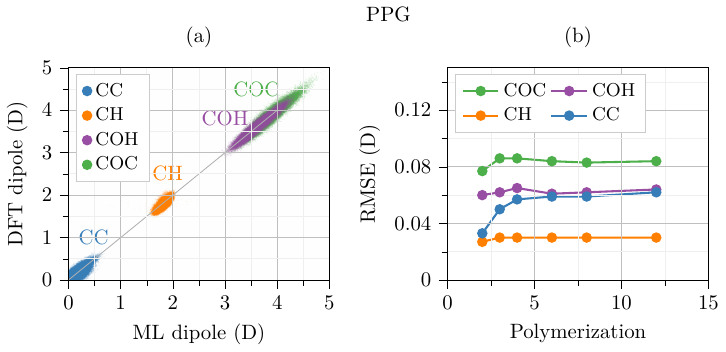}

  \captionsetup[subfigure]{labelformat=empty}
\begin{subcaptionblock}{0.45\linewidth}
  \subcaption{} \label{fig:la_ppg_a} 
\end{subcaptionblock}
 \begin{subcaptionblock}{0.45\linewidth}
  \subcaption{} \label{fig:la_ppg_b} 
\end{subcaptionblock}
 \caption{(a) Learning accuracy of \ac{ML} bond dipoles of PG2, PG3, PG4, PG6, PG8, and PG12. The blue, orange, purple, and green dots represent the absolute values of the CC, CH, COH, and COC bond dipoles, respectively. (b) The chain length dependence of the prediction accuracy in terms of RMSE.}
\label{fig:la_ppg}
 \end{figure*}

 \begin{figure}[htb]  \captionsetup[subfigure]{font={bf,large}, skip=1pt, margin=-0.7cm,justification=raggedright, singlelinecheck=false}
\centering
\begin{subcaptionblock}{\linewidth}
\includegraphics[]{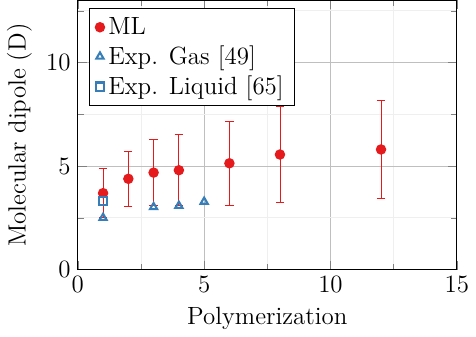}
 \end{subcaptionblock}\hfill
\caption{The average molecular dipole moments of \acp{PPG} with various chain lengthes. For each system, $1000$ molecular dipole moments were randomly sampled from $\SI{200}{\pico\second}$ DPMD calculations of $64$ molecule systems. Experimental values are taken from \cite{ikada1986Liquid} for the gas phase and \cite{vishwam2017Dielectric} for the liquid phase.}
\label{fig:dipole}
\end{figure} \begin{figure}[htb]  \captionsetup[subfigure]{font={bf,large}, skip=1pt, margin=-0.7cm,justification=raggedright, singlelinecheck=false}
\centering
\begin{subcaptionblock}{\linewidth}
\subcaption{} \label{fig:diel_pg2_a}
\includegraphics[width=\textwidth]{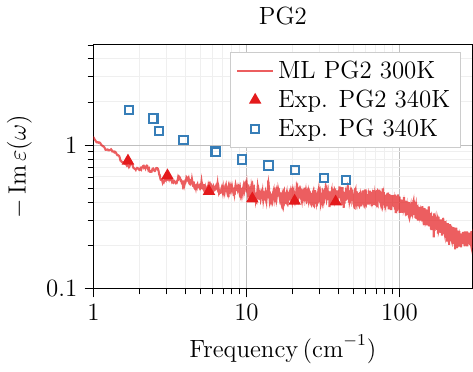}
 \end{subcaptionblock}\hfill
\begin{subcaptionblock}{\linewidth}
\subcaption{} \label{fig:diel_pg2_b}
\includegraphics[width=\textwidth]{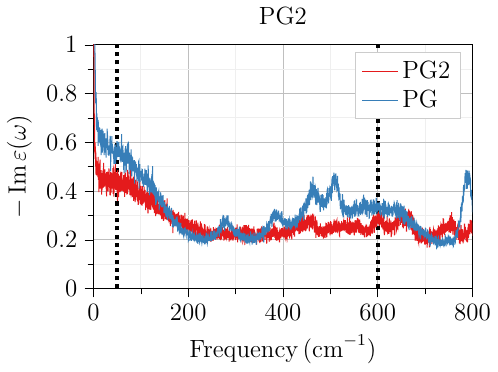}
 \end{subcaptionblock}\hfill
\caption{(a) Calculated dielectric function of PG2 at $\SI{300}{\kelvin}$ (red) togather with the experimental values at $\SI{340}{\kelvin}$ (red triangle)~\cite{kohler2010Glassy}. We also show the exmerimental values of \ac{PG} at $\SI{340}{\kelvin}$ (blue triangle) for comparison. The spectrum was calculated from five independent $\SI{4}{\nano\second}$ trajectories with the moving average method. (b) Calculated dielectric function of PG2 (red) and PG (blue) at $\SI{300}{\kelvin}$. The vertical dotted lines correspond to the peak position of the \ac{VDOS} and libration peak at $50$ and $\SI{600}{\per\cm}$, respectively.} 
\end{figure} \begin{figure}[htb]  \captionsetup[subfigure]{font={bf,large}, skip=1pt, margin=-0.7cm,justification=raggedright, singlelinecheck=false}
\centering
\begin{subcaptionblock}{\linewidth}
\includegraphics[]{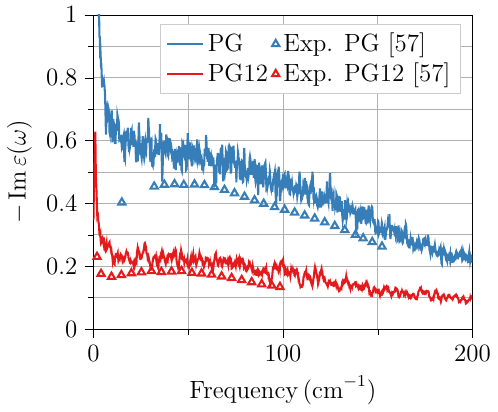}

\end{subcaptionblock}\hfill
\caption{The calculated imaginary part of the dielectric functions of PG and PG12 at $\SI{300}{\kelvin}$ with experimental data for PG and PG12 at $\SI{300}{\kelvin}$~\cite{koda2016Broadband}. The data of PG is the same as that shown in \cref{fig:dielec_pg}.}
\label{fig:dielec_ppg}
\end{figure} 
To estimate the impact of intermolecular interactions on the dielectric function, we decomposed the total dielectric function into intra and intermolecular components. The total dipole moment of the system can be expressed as the sum of the molecular dipoles as $\vb*{M}(t) = \sum_{i}\vb*{\mu}^{\mathrm{mol}}_{i}(t)$, and the dipole autocorrelation function can be calculated as the sum of the intramolecular and intermolecular correlation functions as 
 \begin{align}\expval{\vb{M}(0)\cdot \vb{M}(t)} &= \sum_{i}\expval{\vb*{\mu}^{\mathrm{mol}}_{i}(0)\cdot \vb*{\mu}^{\mathrm{mol}}_{i}(t)} \notag\\
  &+\sum_{i\neq j}\expval{\vb*{\mu}^{\mathrm{mol}}_{i}(0)\cdot \vb*{\mu}^{\mathrm{mol}}_{j}(t)}.\label{195111_29Feb24}
 \end{align}
The total dielectric function can be divided into contributions from intramolecular and intermolecular interactions:
\begin{align}
\varepsilon''(\omega) &=\frac{\omega}{6\varepsilon^{0}k_{\mathrm{B}}TV} \int_{-\infty}^{\infty}\left(\sum_{i}\expval{\vb*{\mu}^{\mathrm{mol}}_{i}(0)\cdot \vb*{\mu}^{\mathrm{mol}}_{i}(t)} \right. \notag \\
 &\left.+\sum_{i\neq j}\expval{\vb*{\mu}^{\mathrm{mol}}_{i}(0)\cdot \vb*{\mu}^{\mathrm{mol}}_{j}(t)}\right) e^{-i\omega t} \dd t \\
 &= \varepsilon''_{\mathrm{intra}}(\omega)+\varepsilon''_{\mathrm{inter}}(\omega). \label{Eq:decompose}
\end{align}

\Cref{fig:dielec_pg_b} shows the dielectric function at $\SI{300}{\kelvin}$, decomposed into intra and intermolecular components, in which we have discovered two key features. First, the intermolecular component sharply increases in the low-frequency region, as observed in water~\cite{holzl2021Dielectric,carlson2020Exploring}. Recent experiments have been revealing that the $\alpha$ process in hydrogen-bonded liquids is imputed to not a single-molecule orientational relaxation, but a relaxation of the hydrogen-bond network involving multiple molecules~\cite{hansen2016Identification,sillren2014Liquid,bohmer2014Structure}, and our calculations suggest that this is also true for \ac{PG}. Second, the intermolecular-component exhibits a small peak around $\SI{100}{\per\cm}$, while the intramolecular-component peaks near $\SI{40}{\per\cm}$. Koda et al.~\cite{koda2016Broadband} experimentally studied the dielectric spectra of liquid \ac{PG} at room temperature, observing a prominent peak at $\SI{50}{\per\cm}$ and a small shoulder at $\SI{120}{\per\cm}$, attributing the latter to intermolecular interactions. Our discovered peak in the intermolecular-component is located at nearly the same frequency as the experimentally observed one, providing theoretical support that this peak is due to correlations of intermolecular dipole moments. For hydrogen-bonded liquid molecules, a broad intermolecular peak in the THz region is generally attributed to hydrogen-bond stretching. In the case of water, the intermolecular part exhibits a significant peak at $\SI{200}{\per\cm}$, the magnitude of which exceeds that of the intramolecular part. The intermolecular peak we found is considerably smaller than that of the intramolecular part, and investigating the cause of this discrepancy remains a subject for future investigation. Additionally, we observe that the libration motion at $\SI{600}{\per\cm}$ is predominantly explained by the intramolecular component, similar to other hydrogen-bonded liquids~\cite{amano2024Chemical,carlson2020Exploring}. This indicates that the libration motion of hydroxyl hydrogen rarely induces dipole moments in neighbor molecules, unlike hydrogen-bond stretching motions.

\Cref{fig:hbond_pg} shows the power spectra of $r_{\mathrm{OO}}$ and $\vb*{e}_{\mathrm{OH}}$, where $r_{\mathrm{OO}}$ represents the shortest intermolecular oxygen-oxygen distance, reflecting hydrogen bond stretching motions, and $\vb*{e}_{\mathrm{OH}}$ denotes the unit direction vector of the OH bonds, corresponding to hydrogen motions. These spectra were calculated from $\SI{2}{\nano\second}$ trajectories at $\SI{240}{\kelvin}$. The $\vb*{e}_{\mathrm{OH}}$ spectrum exhibits a prominent peak around $\SI{600}{\per\cm}$, corresponding to the position of the libration peak. In contrast, the $r_{\mathrm{OO}}$ spectrum shows a significant peak around $\SI{100}{\per\cm}$, which is associated with both the VDOS peak and the intermolecular interaction peak. Therefore, intermolecular hydrogen bonding plays a major role in the VDOS peak, even though it is not distinctly visible in the \ac{VDOS} spectra in \cref{fig:vdosa}. 
 
\subsection{Oligomers}

We examined the transferability of the chemical bond-based \ac{ML} model using \acp{PPG} of varying chain lengths. We trained the \ac{ML} dipole models using PG2, which, unlike PG, contains a COC segment (ether bond) and provides ideal training data for broader applications across PPGs. The model trained on PG2 was then applied to PG3, PG4, PG6, PG8, and PG12 to assess its accuracy and dielectric properties. \Cref{table:rmse} summarizes the \ac{RMSE} of PG2 models applied to PG2 validation data, confirming that the prediction accuracy is almost identical to that of the PG model. The \ce{COH} and \ce{COC} models provide slightly better accuracy than the sum of the \ce{CO}, lone pair, and \ce{OH} models, confirming that our new approach is also effective for PG2.

\Cref{fig:la_ppg_a} illustrates the robust predictive performance of the PG2 model in estimating bond dipoles across PG oligomers. For validation, we randomly selected $1000$ data points from each PG3, PG4, PG6, PG8, and PG12 and compared the \ac{ML} predictions to reference DFT calculations. The magnitude of each bond dipole remained nearly consistent, regardless of chain length. \Cref{fig:la_ppg_b} shows the prediction accuracy as \ac{RMSE} for each chain length. The model exhibited the highest accuracy for $n=2$, with only a slight decline in precision as $n$ increased. The accuracy is marginally lower for COC and CC bonds than for CH and COH bonds, which is likely due to electron delocalization as the chain length increases. These results corroborate that the PG2-trained model retains its accuracy and transferability, even when applied to various \acp{PPG} not referenced in the training.

\Cref{fig:dipole} compares calculated molecular dipoles in the liquid phase at several chain lengths with experimental values of liquids at $\SI{298}{\kelvin}$~\cite{vishwam2017Dielectric} and gases at $\SI{293}{\kelvin}$~\cite{ikada1979Dielectric}. Error bars represent the standard deviations. The computed average dipole moment of \ac{PG} was $\SI{3.6}{\mathrm{D}}$, which closely agreed with the experimental value of $\SI{3.2}{\mathrm{D}}$~\cite{vishwam2017Dielectric}. The calculated molecular dipoles in the liquid phase were about $\SI{1}{\mathrm{D}}$ larger than the experimental values of the gas phase, which is a well-known phenomenon for hydrogen-bonded molecules~\cite{jorge2021Selfconsistent,amano2024Chemical}. Our analysis further reveals that, similarly to experimental observations in the gas phase, the dipole moment in the liquid phase also increases slightly with chain length. Given that the COH bond dipole does not significantly change with chain length, the increase in this dipole is likely caused by fluctuations in the dipole of the alkyl groups in either the main or side chains.

\Cref{fig:diel_pg2_a} demonstrates that the predicted dielectric function of PG2 at $\SI{300}{\kelvin}$ closely matches the experimental value at $\SI{340}{\kelvin}$~\cite{kohler2010Glassy}, confirming the reliability of the trained ML models. The intensity of the orientation relaxation of PG2 is clearly smaller than that of \ac{PG}. The observed temperature discrepancy with the experiment is consistent with that in \cref{fig:dielec_pg_a} for \ac{PG}, suggesting that approximations such as the exchange-correlation functional and neglecting quantum effects may influence the relaxation dynamics of \ac{PG} and PG2 similarly. \Cref{fig:diel_pg2_b} compares the THz dielectric function of \ac{PG} and PG2. The peak intensity of the libration of PG2 at $\SI{600}{\per\cm}$ is smaller than that of PG and is buried in the narrow peaks of local vibrational modes. This is because the number of \ce{OH} groups in PG2 is half that of PG, reducing the generated dipole moments of the libration. On the other hand, the \ac{VDOS} peak of PG2 is only slightly smaller than that of \ac{PG} and has a very similar shape. The VDOS peak, therefore, originates not only from \ce{OH} groups but also from other atoms. \Cref{fig:dielec_ppg} shows that the \ac{VDOS} peak of the dielectric function of PG12, using the \ac{ML} potential and dipole moments trained on \ac{PG2} training data, agrees well with experimental values~\cite{koda2016Broadband}. This confirms that the \ac{PG2} \ac{ML} models can reliably predict the dielectric properties of \acp{PPG} with longer chain lengths, which are not referenced during training. The spectrum was computed from five independent $\SI{2}{\nano\second}$ trajectories with $64$ molecules, in a total of $7872$ atoms. Notably, the peak position is almost identical regardless of the chain length, while the peak intensity decreases with increasing chain length. The decrease in intensity may be due to the reduction in the density of the \ce{OH} groups with increasing chain length.

\section{Conclusion}\label{sec:conclusion}

Through the study of the dielectric properties of \ac{PG} and \ac{PPG}, we validated the transferability and effectiveness of our bond-based \ac{ML} dipole moment scheme combined with \ac{ML} potentials. In this scheme, we assigned \acp{WC} to each chemical bond, and trained \ac{ML} models to predict the \ac{WC} for each bond type. We adopted a slightly modified approach from the previous study to improve accuracy while reducing computational costs, which is suitable for long-time and large-scale calculations. Furthermore, to achieve extended simulation times of up to $\SI{20}{\nano\second}$, the ML potentials were trained using active sampling.

We first investigated the dielectric properties of \ac{PG}, which is the first detailed analysis from first-principles to the best of our knowledge. Our developed \ac{ML} dipole models demonstrated high accuracy, and successfully reproduced the dielectric function in both the THz and GHz regions. Notably, the GHz dielectric function was calculated using $\SI{20}{\nano\second}$ \ac{MD} trajectories, a time scale beyond the reach of \ac{AIMD}. In the THz region, the dominant features are the \ac{VDOS} peak at $\SI{50}{\per\cm}$ and the hydroxyl hydrogen libration peak at $\SI{600}{\per\cm}$. The \ac{VDOS} peak reflects overall molecular vibrations involving carbon, hydrogen, and oxygen atoms and is attributed to intramolecular components of the dielectric function. We also found a shoulder at $\SI{100}{\per\cm}$ in the dielectric function of the intermolecular component, which was previously found experimentally. Our analysis of the oxygen-oxygen interatomic power spectrum suggests that the hydrogen bond stretching may contribute significantly to the \ac{VDOS} peak. In the GHz range, the intermolecular components become dominant at lower frequencies below $\SI{0.4}{\per\cm}$.

We then applied the \ac{ML} dipole models, trained on PG2 data, to \acp{PPG} with longer chain lengths which were not included in the training data. The models accurately predicted the bond dipoles of \acp{PPG} regardless of the chain length. The molecular dipole moment of liquid \acp{PPG} gradually increases with chain length. Furthermore, we calculated the dielectric function of PG12, demonstrating good agreement with experimental data. Considering that the structure of PG12 was not included in the training data, our calculation underscores the high transferability of our approach. We expect the current approach to be the first step toward developing a universal bond dipole model for various molecular systems.

\begin{acknowledgments}
This research was funded by a JST-Mirai Program Grant Number JPMJMI20A1, a MEXT Quantum Leap Flagship Program (MEXT Q-LEAP) grant number JPMXS0118067246, Japan, and JSR Corporation via JSR-UTokyo Collaboration Hub, CURIE. The computations in this study have been conducted using computational resources of the supercomputer Fugaku provided by the RIKEN Center for Computational Science (ProjectID: hp220331, hp230124, and hp240148) and the facility of the Supercomputer Center, the Institute for Solid State Physics, the University of Tokyo.
\end{acknowledgments}

\appendix
\begin{figure}[htb]\centering
\begin{subcaptionblock}{0.45\linewidth}
  \subcaption{$\SI{498}{\per\cm}$} \centering
  \adjustbox{right}{\includegraphics[width=\textwidth]{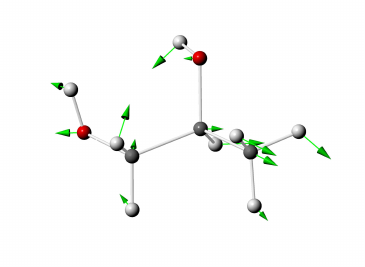}}
 \end{subcaptionblock}
 \begin{subcaptionblock}{0.45\linewidth}
  \subcaption{$\SI{456}{\per\cm}$} \centering
  \adjustbox{left}{\includegraphics[width=\textwidth]{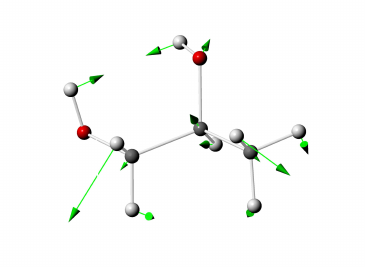}}
 \end{subcaptionblock}
 \begin{subcaptionblock}{0.45\linewidth}
  \subcaption{$\SI{416}{\per\cm}$} \centering
  \adjustbox{right}{\includegraphics[width=\textwidth]{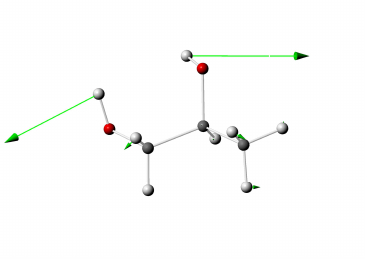}}
 \end{subcaptionblock}
 \begin{subcaptionblock}{0.45\linewidth}
  \subcaption{$\SI{356}{\per\cm}$} \centering
  \adjustbox{left}{\includegraphics[width=\textwidth]{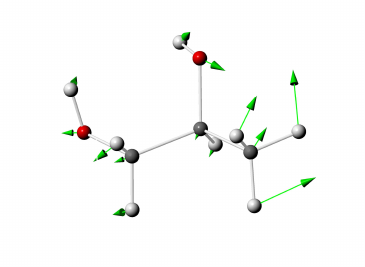}}
 \end{subcaptionblock}
 \begin{subcaptionblock}{0.45\linewidth}
  \subcaption{$\SI{238}{\per\cm}$} \centering
  \adjustbox{right}{\includegraphics[width=\textwidth]{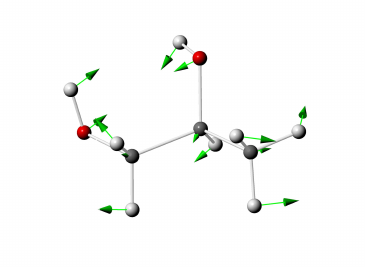}}
 \end{subcaptionblock}
 \begin{subcaptionblock}{0.45\linewidth}
  \subcaption{$\SI{204}{\per\cm}$} \centering
  \adjustbox{left}{\includegraphics[width=\textwidth]{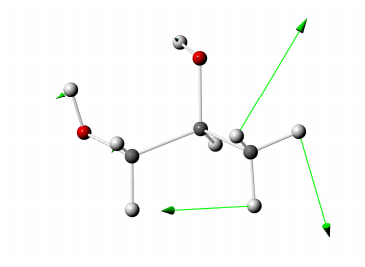}}
 \end{subcaptionblock}
 \caption{Calculated vibrational modes of the gas phase PG, the frequencies of which range from $200$ to $\SI{500}{\per\cm}$. Oxygen, carbon, and hydrogen are represented by red, gray, and white spheres, respectively. The green arrows represent the magnitude and direction of the eigenvectors of the vibration modes.}
\label{fig:vib_analysis}
 \end{figure}

\section{Vibrational analysis of gas phase \ac{PG}}\label{appendix:A}
Here, we study the origin of several sharp peaks in \ac{VDOS} of \cref{fig:vdosa} at $280$, $390$, $460$, and $\SI{520}{\per\cm}$, which show contributions from all C, H, and O atoms. These peaks are local vibrational modes and can be analyzed using gas phase calculations~\cite{koda2016Broadband}. We employed VASP~\cite{kresse1996Efficient,kresse1999Ultrasoft} for structual relaxation and finite difference vibrational analysis with the BLYP functional. The plane-wave energy cutoff was set to $\SI{600}{\electronvolt}$, with an energy convergence threshold of $\SI{1e-9}{\electronvolt}$. We placed one \ac{PG} molecule in the cubic cell with the lattice parameter of $\SI{20}{\angstrom}$, with only the $\Gamma$ point used for electronic integration. \Cref{fig:vib_analysis} shows the six calculated vibrational modes, the frequencies of which range from $200$ to $\SI{600}{\per\cm}$. Among these, the $\SI{204}{\per\cm}$ vibration corresponds to the rotational motion of alkyl hydrogens, and the $\SI{416}{\per\cm}$ vibration is attributed to the librational motion of hydroxyl hydrogens, which do not appear in \cref{fig:vdosa} as these two modes only have significant contribution from hydrogen atoms. Therefore, the remaining four modes, representing vibrational motions of the main molecular chain, correspond to sharp peaks observed in \cref{fig:vdosa}.

 
%
%
\end{document}